\documentstyle[preprint,prl,aps]{revtex}

\newif\ifLONG       
\LONGfalse         

\begin{document}
\draft
\title{A Four Dimensional Generalization of the Quantum Hall Effect}
\author{Shou-Cheng Zhang, Jiangping Hu}
\address{
Department of Physics, Stanford University, Stanford, CA 94305
}
\address{
Center for Advanced Study, Tsinghua University, Beijing, China}


\maketitle
\begin{abstract}
We construct a
generalization of the quantum Hall effect, where particles move in
four dimensional space under a $SU(2)$ gauge field.
This system has a macroscopic number of degenerate single particle
states. At appropriate integer or fractional filling fractions the system forms
an incompressible quantum liquid.
Gapped elementary excitations in the bulk interior and gapless elementary
excitations at the boundary are investigated.

\end{abstract}
\newpage

\narrowtext
Most strongly correlated systems develop long range order in
the ground state. Familiar ordered states include superfluidity,
superconductivity, antiferromagnetism and charge density wave\cite{anderson}.
However, there are special quantum disordered ground
states with fractionalized elementary excitations. In one dimensional
systems, Bethe's Ansatz\cite{bethe} gives exact ground state
wave functions of a class of Hamiltonians,
and the elementary excitations are fractionalized objects
called spinons and holons. In two dimensional quantum Hall
effect (QHE)\cite{laughlin,girvin}, Laughlin's wave function\cite{laughlin}
describes an incompressible
quantum fluid with fractionally charged elementary excitations.
This incompressible liquid can also be described by a
Chern-Simons-Landau-Ginzburg field theory\cite{cslg}, whose long distance limit
depends only on the topology, but not on the metric of the underlying space\cite{witten}.
These two special quantum disordered ground states are the focus of much theoretical
and experimental studies, since they give deep insights on the interplay between
quantum correlations and dimensionality, and on how
this interplay can give rise to fractionalized elementary
excitations.

In view of their importance, it is certainly
desirable to generalize these quantum wave functions to higher
dimensions. However, despite repeated efforts, the
Bethe's Ansatz solutions have not yet been generalized to dimensions
higher than one. Laughlin's wave function uses properties which
seem to be special to the two dimensional space.
In this work we shall report the generalization
of the quantum Hall system to four space dimensions, and this
system shares many compelling similarities to the two dimensional
counterpart.
In the two dimensional (2D) QHE, the charge current
is carried in a direction perpendicular to the applied electric field (and
also perpendicular to the magnetic field, which is applied normal to the
2D electron gas).
In four space dimensions (4D), there are three independent directions
normal to the electric field, and there appears to be no unique direction
for the current. A crucial ingredient of our generalization is that the particles
also carry an internal $SU(2)$ spin degree of freedom. Since there are exactly three
independent directions for the spin, the particle current can be uniquely
carried in the direction where the spins point. At special filling factors,
the quantum disordered ground state of our 4D QHE is separated from all excited states by a
finite energy gap, and the lowest energy excitations are fractionally charged
quasi-particles.

While all excitations have finite energy gaps in the bulk interior, elementary
excitations at the three dimensional boundary of this quantum fluid are gapless, in analogy
with the edge states of the quantum Hall effect\cite{halperin,wen,stone}.
These boundary excitations could be used to model the relativistic elementary
particles, such as photons and gravitons. In contrast to conventional quantum
field theory approach, this model has the advantage that the short distance
physics is finite and self-consistent. In fact, the magnetic length in this
model provides a fundamental lower limit on all length scales. This feature
shares similarity to non-commutative quantum field theory and string theory
of elementary particles.

{\bf A four dimensional generalization of the quantum Hall problem}
In the QHE problem, it is advantageous to consider compact spherical
spaces which can be mapped to the flat Euclidean spaces by the standard
stereographical mapping\cite{haldane}. Eigenstates in
the QHE problem are called Landau levels, and we first review the
lowest Landau level (lll) defined on the 2D sphere, denoted
by $S^2$. A point $X_i$ on $S^2$ with radius $R$
can be described by dimensionless vector coordinates $x_i=X_i/R$, with
$i=1,2,3$, which satisfy
$x_i^2=1$. However, $S^2$ has a special property that one can also
take the ``square root" of the vector coordinate $x_i$ through the
introduction of the complex spinor coordinates $\phi_\sigma$, with
$\sigma=1,2$. These spinor coordinates are defined by
\begin{eqnarray}
x_i = \bar\phi_\sigma (\sigma_i)_{\sigma\sigma'} \phi_{\sigma'}
\ \ \, \ \ \
\bar\phi_\sigma \phi_\sigma =1
\label{1stHopf}
\end{eqnarray}
where $\sigma_i$ are the three Pauli spin matrices. If there is
a magnetic monopole of strength $g$ at the center of $S^2$, satisfying
the Dirac quantization condition $eg=I$=half integer, then the
normalized eigenfunctions in the lll are just the algebraic products of the
spinor coordinates
\begin{eqnarray}
\langle x|I,m \rangle =
\sqrt{\frac{(2I)!}{(I+m)!(I-m)!}}\phi_1^{I+m}\phi_2^{I-m}
\label{spinstate}
\end{eqnarray}
Here $m=-I,-I+1,...I-1,I$, therefore the ground state is $2I+1$
fold degenerate. Any states in the lll can be expanded in terms of
a homogeneous
polynomial of $\phi_1$ and $\phi_2$ with degree $2I$. Notice that the
conjugate coordinate $\bar\phi_\sigma$ does not enter the wave
function in the lll.

We see that the crucial algebraic structure of the QHE
problem is the fractionalization of a vector coordinate into
two spinor coordinates. Therefore, in seeking a higher dimensional
generalization of the QHE problem we need to find
a proper generalization of Eq. \ref{1stHopf}. As
the generalization of the three Pauli matrices are the
five $4\times 4$ Dirac matrices $\Gamma_a$, satisfying the
Clifford algebra $\{\Gamma_a,\Gamma_b\}=2\delta_{ab}$, we
generalize Eq. \ref{1stHopf} to
\begin{eqnarray}
x_a = \bar\Psi_\alpha (\Gamma_a)_{\alpha\alpha'} \Psi_{\alpha'}
\ \ \, \ \ \
\bar\Psi_\alpha \Psi_\alpha =1
\label{2ndHopf}
\end{eqnarray}
Here, $\Psi_\alpha$ is a four component complex spinor with $\alpha=1,2,3,4$,
and $x_a$ is a five component real vector. From the normalization condition of the
$\Psi$ spinor it may be seen that $x_a^2=1$, therefore,
$X_a=Rx_a$ describes a point of the 4D sphere $S^4$
with radius $R$.
From this heuristic reasoning one may hope to find a four
dimensional generalization of the QHE problem, where the
wave functions in the ground states are described by the products
of $\Psi_\alpha$ spinors, in a natural generalization of Eq. (\ref{spinstate}).
Eq. (\ref{1stHopf}) and Eq. (\ref{2ndHopf}) are known in the
mathematical literature as the first and the second Hopf maps\cite{holonomy}.
The problem now is to find a Hamiltonian for
which these are the exact ground state wave functions.

An explicit solution to the Eq. (\ref{2ndHopf}) can be
expressed as
\begin{eqnarray}
& & \Gamma^{(1,2,3)} \!=\! \left( \begin{array}{cc}
               0   & i \sigma_i  \\
               -i \sigma_i       &  0  \end{array} \right) \ \ , \ \
 \Gamma^4\!=\! \left( \begin{array}{cc}
               0          & 1  \\
               1          & 0            \end{array} \right) \ \ , \ \
 \Gamma^5 \!=\! \left( \begin{array}{cc}
                1         & 0  \\
                0         & -1            \end{array} \right) \\
& &  \left( \begin{array}{c}
               \Psi_1    \\
               \Psi_2  \end{array} \right) =
               \sqrt{\frac{1+x_5}{2}}
               \left( \begin{array}{c}
               u_1    \\
               u_2  \end{array} \right) \ \ , \ \
\left( \begin{array}{c}
               \Psi_3    \\
               \Psi_4  \end{array} \right) =
               \sqrt{\frac{1}{2(1+x_5)}} (x_4-ix_i \sigma_i)
               \left( \begin{array}{c}
               u_1    \\
               u_2  \end{array} \right)
\label{explicit}
\end{eqnarray}
where $(u_1,u_2)$ is an arbitrary two component complex spinor
satisfying $\bar u_\sigma u_\sigma=1$. Any $SU(2)$ rotation on
$u_\sigma$ preserves the normalization condition, and maps to the
same point $x_a$ on $S^4$. From the explicit form of
$\Psi_\alpha$, one can compute the geometric connection (Berry's phase)
$\bar\Psi_\alpha d\Psi_\alpha$\cite{holonomy}, where the
differentiation operator $d$ acts on the vector coordinates $x_a$,
subject to the condition $x_a d x_a=0$. One finds
$\bar\Psi_\alpha d\Psi_\alpha = \bar u_\sigma (a_a
dx_a)_{\sigma\sigma'} u_{\sigma'}$, $a_5=0$, and
\begin{eqnarray}
a_\mu = \frac{-i}{1+x_5}\eta_{\mu\nu}^i x_\nu I_i \ \ , \ \
\eta_{\mu\nu}^i=\epsilon_{i\mu\nu4}+\delta_{i\mu}\delta_{4\nu}-
\delta_{i\nu}\delta_{4\mu}
\label{potential}
\end{eqnarray}
where $I_i=\sigma_i/2$ and $\eta_{\mu\nu}^i$ is also known as
the t'Hooft symbol. $a_\mu$ is the $SU(2)$ gauge potential of a
Yang monopole defined on $S^4$\cite{yang1}. Upon a conformal transformation
from $S^4$ to the 4D Euclidean space
$R^4$\cite{jackiw}, this gauge potential is transformed to the
instanton solution of the $SU(2)$ Yang-Mills
theory\cite{instanton}. We shall call $I_i$ a $SU(2)$ isospin
matrix, and the gauge potential defined in Eq. \ref{potential} can
be generalized to an arbitrary representation $I$ of the
$SU(2)$ Lie algebra $[I_i,I_j]=i\epsilon_{ijk} I_k$. The gauge
field strength can be calculated from the form
of the gauge potential. From the covariant derivative
$D_a=\partial_a+a_a$, we define the field strength as
$f_{ab}=[D_a,D_b]$. Both $a_a$ and $f_{ab}$ are matrix valued, and
can be generally expressed in terms of the isospin components
$a_a=-ia_a^i I_i$ and $f_{ab}=-if_{ab}^i I_i$. In terms of these
components, we find $f_{5\mu}^i = - (1+x_5) a_\mu^i$ and
$f_{\mu\nu}^i = x_\nu a_\mu^i - x_\mu a_\nu^i - \eta_{\mu\nu}^i$.
In addition to the dimensionless quantities $a_\mu$ and $f_{ab}$,
we shall sometimes also use dimensionful quantities defined by
$A_\mu = R^{-1} a_\mu (X/R)$, and $F_{ab}= R^{-2}f_{ab}(X/R)$.

With this introduction and motivation, we are now in a position to
introduce the Hamiltonian of our quantum mechanics problem.
The symmetry group of $S^4$ is $SO(5)$, generated by the angular
momentum operator $L^{(0)}_{ab}= -i(x_a \partial_b - x_b \partial_a)$.
The Hamiltonian of a single particle moving on $S^4$ can be
expressed as $H=\frac{\hbar^2}{2MR^2}\sum_{a<b}(L_{ab}^{(0)})^2$, where $M$ is
the inertia mass, and $R$ is the radius of $S^4$.
Coupling to a gauge field $a_a$ may be introduced by replacing
$\partial_a$ with the covariant derivative $D_a$.
Under this replacement,
$L_{ab}^{(0)}$ becomes $\Lambda_{ab}= -i(x_a D_b - x_b D_a)$. The
Hamiltonian of our generalized QHE problem is therefore
given by
\begin{eqnarray}
H=\frac{\hbar^2}{2MR^2}\sum_{a<b}\Lambda_{ab}^2
\label{H}
\end{eqnarray}
This Hamiltonian has an important parameter $I$, defined by
$I_i^2=I(I+1)$, which specifies the dimension of the $SU(2)$
representation in the potential (Eq. \ref{potential}).

Unlike $L_{ab}^{(0)}$, $\Lambda_{ab}$ does not satisfy the $SO(5)$
commutation relation. However, one can define
$L_{ab}=\Lambda_{ab}-if_{ab}$, which does satisfy the $SO(5)$ commutation relation.
While only a subset of $SO(5)$ irreducible representations can
be generated from the $L_{ab}^{(0)}$ operators, Yang\cite{yang2} showed that
$L_{ab}$ generates all $SO(5)$ irreducible representations.
In general, a $SO(5)$ irreducible representation is labeled
by two integers $(p,q)$, with $p\ge q\ge 0$. For such a representation, the
Casimir operator and the dimensionality are given by
$C(p,q)=\sum_{a<b}L_{ab}^2=\frac{p^2}{2}+\frac{q^2}{2}+2p+q$ and
$d(p,q)=(1+q)(1+p-q)(1+\frac{p}{2})(1+\frac{p+q}{3})$
respectively. However, for a given $I$, these
two integers are related by $p=2I+q$. One can show that
$\sum_{a<b}\Lambda_{ab}^2 = \sum_{a<b}L_{ab}^2 - 2 I_i^2$.
Therefore, for a given $I$, the energy eigenvalues of
the Hamiltonian (Eq. \ref{H}) is given by
\begin{eqnarray}
E(p=2I+q,q)=\frac{\hbar^2}{2MR^2} (C(p=2I+q,q)-2I(I+1))
\label{E}
\end{eqnarray}
with degeneracy $d(p=2I+q,q)$. The ground state, which is the
lowest $SO(5)$ level for a given $I$, is obtained by setting
$q=0$, and we see that it is $\frac{1}{6}(p+1)(p+2)(p+3)$ fold
degenerate. Therefore, the dimension of the $SU(2)$ representation
plays the role of the magnetic flux, while
$q$ plays the role of the Landau level index. States
with $q>0$ are separated from the ground state by a finite
energy gap.

Besides the energy eigenvalues and the degeneracy, we need to know
the explicit form of the ground state wave function.
Yang\cite{yang2} did find the wave function for all the $(p,q)$
states, however, his solution is expressed in a basis that is hard
to work with for our purpose. Realizing the spinor structure we
outlined above, we can express the wave functions of the lowest
$SO(5)$ levels $(p,0)$ in a very simple form. First, one can check
explicitly that $\Psi_\alpha$ given in Eq. \ref{explicit}
is indeed an eigenfunction of the Hamiltonian (Eq. \ref{H}) with
$I=1/2$. This follows from the fact that it is a $SO(5)$ spinor
under the generators $L_{ab}$: $L_{ab} \Psi_\alpha = - \frac{1}{2}
(\Gamma_{ab})_{\alpha\beta} \Psi_\beta$. From this one can see
that $\Psi_{\alpha_1...\alpha_p}(x)=\Psi_{\alpha_1} \cdot \cdot \cdot
\Psi_{\alpha_p} $ transforms as an irreducible spinor under the
$SO(5)$ group. Therefore, the complete set of normalized basis
functions in the lowest $SO(5)$ level $(p,0)$ with orbital
coordinate $x_a=\bar\Psi \Gamma_a \Psi$ and isospin coordinate
$n_i=\bar u \sigma_i u$ is given by
\begin{eqnarray}
\langle x_a,n_i|m_1,m_2,m_3,m_4\rangle = \sqrt{\frac{p!}{m_1! m_2! m_3! m_4!}}
\Psi_1^{m_1} \Psi_2^{m_2} \Psi_3^{m_3} \Psi_4^{m_4}
\label{wavefunction}
\end{eqnarray}
with integers $m_1+m_2+m_3+m_4=p$. This set of basis functions in
the lowest $SO(5)$ level are the exact eigenstates of the
Hamiltonian (Eq. \ref{H}) with $\frac{1}{6}(p+1)(p+2)(p+3)$ fold
degenerate eigenvalue of $\frac{\hbar^2}{2MR^2}p$. They are the
natural generalizations of the wave functions in the lll
(Eq. \ref{spinstate}) of the QHE problem. The very simple
form of the single particle wave function (Eq. \ref{wavefunction})
introduced here greatly helps calculations of the many-body wave
function.

{\bf An incompressible quantum spin liquid}
We are now in the position to consider the quantum many body
problem involving $N$ fermions. The simplest case to consider
is $N=d(p,0)$, when the lowest $SO(5)$ level is completely
filled. In this case, the filling factor $\nu\equiv N/d(p,0)=1$,
and the many-body ground state wave function
is unique.

Before presenting the explicit form of the wave function, we first
need to discuss the thermodynamic limit in this problem, as it
is rather non-trivial. We shall consider the limit
$p=2I\rightarrow\infty$ and $R\rightarrow\infty$ while keeping $q$
constant. For energy eigenvalues in Eq. \ref{E} to be finite, we
need $l_0=\lim_{R\rightarrow\infty} \frac{R}{\sqrt{p}}$ to
approach a finite constant, which can be defined as the ``magnetic
length" in this problem. In this limit,
$E(q)=\frac{\hbar^2}{2Ml_0^2}(1+q)$ and the single particle energy
spacing is finite. At $\nu=1$, $N\sim p^3\sim R^6$, the naively
defined particle density $N/R^4$ would be infinite. However, we
need to keep in mind that each particle also have an infinite
number of isospin degrees since $I\rightarrow\infty$. Taking this
fact into account, we see that the volume of the configuration
space, defined to be the product of the volume in orbital and
isospin space is $R^4\times R^2$. Therefore, the density $n=N/R^6$
is actually finite in this limit.

Using $A=\{m_1,m_2,m_3,m_4\}=1,..,d(p,0)$ to label the single
particle states, the many particle wave function is given by
a Slater determinant.
\begin{eqnarray}
\Phi(x_1,...,x_N)=\Psi_{A_1}(x_1) \cdot \cdot \cdot \Psi_{A_N}(x_N)
\epsilon_{A_1...A_N}
\label{Slater}
\end{eqnarray}
The density correlation function
$\rho(x,x')=\frac{1}{(N-2)!}\int dx_3\cdot \cdot dx_N |\Phi(x,x',x_3,..,x_N)|^2$ can
be computed exactly and is given by
\begin{eqnarray}
\rho(x,x')=1-|\bar\Psi_A(x)\Psi_A(x')|^2=1-|\bar\Psi_\alpha(x)\Psi_\alpha(x')|^{2p}
\approx 1-e^{-\frac{1}{4l_0^2}(X_\mu^2+N_\alpha^2)}
\label{rho_explicit}
\end{eqnarray}
where the explicit form of the single particle wave function
(Eq. \ref{wavefunction}) was used. In the approximation, we placed
particle $x'$ on the north poles of both the orbital and
the isospin space, {\it i.e.} $x'_a=\delta_{5a}$ and
$n'_i=\delta_{3i}$, and expanded in terms of
$X_\mu^2=R^2(x_1^2+x_2^2+x_3^2+x_4^2)$ and
$N_\alpha^2=R^2(n_1^2+n_2^2)$ in the limit
$l_0^2=\lim_{R\rightarrow\infty} \frac{R^2}{p}$.
We see that just
like in the QHE liquid, a particle is accompanied by a
perfect correlation hole, gaussianly localized in its vicinity.
The new feature in our case is that the incompressibility applies
to both the charge and isospin channel.

Having discussed the generalization to the integer QHE, let us
now turn to the fractional QHE. One
can see that the many body wave function $\Phi_m=
\Phi^m(x_1,...,x_N)$ with odd integer $m$ is also a legitimate
fermionic wave function in the lowest $SO(5)$ level. This is so
because the product of the basic spinors $\Psi_\alpha$ is always a
legitimate state in the lowest $SO(5)$ level. $\Phi_m$ is a
homogeneous polynomial of $\Psi_\alpha(x_i)$ with degree $p'=mp$.
Therefore, the degeneracy of the lowest $SO(5)$ level in this case
is $d(mp,0)=\frac{1}{6}(mp+1)(mp+2)(mp+3)\rightarrow
\frac{1}{6}m^3 p^3$, while the particle number is still
$N=d(p,0)$. The filling factor in this case is
$\nu=N/d(mp,0)=m^{-3}$. While $\Phi_m$ can not be expressed in the
Laughlin form of a single product, we can still use plasma analogy
to understand its basic physics. $|\Phi_m|^2$ can also be
interpreted as the Boltzmann weight for a classical fluid, whose
effective inverse temperature is $\beta_m=m\beta_{m=1}$. As the
correlation functions for $m=1$ case can be computed exactly, it
is plausible that the $m>1$ case has similar correlations, in
particular, it is also an incompressible liquid. However, the
effective parameters need to be rescaled properly in the
fractional case. The effective magnetic length is given by
$l'_0=\frac{R}{\sqrt{p'}}=\frac{R}{\sqrt{mp}}$. This
incompressible liquid supports fractionalized charge excitation
with charge $m^{-3}$. Such a state may be described by a wave
function of the form $\Phi^{m-1} \Phi_h$, where $\Phi_h$ is the
wave function of the integer case, where one hole is removed from
a given location in the bulk interior to the edge of the fluid. To
our knowledge, this is the first time where a quantum liquid with
fractionaly charge excitation has been identified in higher
dimension $d>2$.

{\bf Emergence of relativity at the edge}
Before we go to the
discussion of our model, let us first review how $1+1$ dimensional
relativity emerges at the edge of the 2D QHE problem. We shall
restrict ourselves to the integer case only.
In the lll, there is no kinetic energy. The only energy is
supplied by the confining potential $V(r)$, which confines the
particles in a circular droplet of size $R$.
Eigenfunctions in the lll takes the form
$\phi_n (z) = z^n exp(-\frac{|z|^2}{4l_0^2})$. From this we
see that a particle is localized in the radial direction at $r_n=n l_0$,
and it carries angular momentum $L=n$. Edge excitations are
particle hole excitations of the droplet. A particle hole
pair with the lll label $n$ and $m$ near the edge has
energy $E=V_n-V_m=(n-m)l_0 V'(R)$,
and angular momentum $L=n-m$. Therefore, a relativistic, linear relationship
exists between the energy and the momentum of the edge excitation.
Furthermore, since $n-m>0$, the edge waves propagate only in one
direction, {\i.e.} they are chiral.
Therefore, we see that relativity emerges at the edge because of
a special relationship between the radial and the angular part of the
wave function $z^n$. It turns out that such a relationship also
exists in the present context.

In our spherical model, we can introduce a confining potential
$V(X_a)=V(X_5)$, where $V(X_5)$ is a monotonic function with a minimum at
the north pole $x_5=1$ and a maximum at $x_5=-1$. For $N<d(p,0)$, the
quantum fluid fills the configuration space around the north pole $x_5=1$,
up to the ``fermi latitude" at $x_5^F$.
Within the lowest $SO(5)$ level, there is no kinetic energy, only
the confining potential $V(x_5)$ determines the energy scale of the
problem.  While the $SO(5)$ symmetry of
the $S^4$ sphere is broken explicitly by the confining potential, the
$SO(4)$ symmetry is still valid. Without loss of generality, we can fill
the orbital and isospin space so that the ground state is a $SO(4)$
singlet.

The orbital $SO(4)$ symmetry is defined to be the rotation in the
$(x_1,x_2,x_3,x_4)$ subspace, generated by the angular
momentum operators $L^{(0)}_{\mu\nu}= -i(x_\mu \partial_\nu - x_\nu \partial_\mu)$
where $\mu,\nu=1,2,3,4$. These angular momentum operators satisfy $SO(4)$
commutation relations, which can be decomposed into the following two
sets of $SU(2)$ angular momentum operators
$K_{1i}^{(0)}=\frac{1}{2}(L_i+P_i)$
and $K_{2i}^{(0)}=\frac{1}{2}(L_i-P_i)$,
where $L_i=\frac{1}{2}\epsilon_{ijk}L_{jk}^{(0)}$, $P_i=L_{4i}^{(0)}$.
Because of the coupling to the Yang monopole
gauge potential, these orbital $SO(4)$ generators are modified into
$K_{1i}=K_{1i}^{(0)}$ and $K_{2i}=K_{2i}^{(0)}+I_i$.
Therefore, all edge states can be classified by their $SO(4)$ quantum
numbers $(k_1,k_2)$, where $K_{1i}^2=k_1(k_1+1)$ and
$K_{2i}^2=k_2(k_2+1)$ respectively.
Applying these operators to the states in the lowest $SO(5)$
level (Eq. \ref{wavefunction}), we find that the state $|m_1,m_2,m_3,m_4\rangle$
has quantum numbers $m_1+m_2=2k_2$, $m_1-m_2=2k_{2z}$, $m_3+m_4=2k_1$
and $m_3-m_4=2k_{1z}$. In particular, the elementary $SO(5)$ spinors
defined in Eq. \ref{explicit} transform according to the $(0,1/2)$ and
$(1/2,0)$ representations of $SO(4)$.

In the subspace of lowest $SO(5)$ levels defined by Eq. \ref{wavefunction},
the orbital coordinate operators $x_a$ can be represented by
$x_a=\frac{1}{p}\Psi\Gamma_a \frac{\partial}{\partial\Psi}$.
From this we see that
the $|m_1,m_2,m_3,m_4\rangle$ state is also an eigenstate of $p x_5$,
which takes quantized values $p x_5=m_1+m_2-m_3-m_4$. Since $m_1+m_2+m_3+m_4=p$,
$\frac{p x_5}{2}$ can range over $p+1$ values:
$-\frac{p}{2},-\frac{p}{2}+1,...,\frac{p}{2}$. Therefore, for a given
$p$, and at a fixed latitude on the orbital space $x_5$, the $SO(4)$
quantum numbers $(k_1,k_2)$ are given by
$2k_1=\frac{p}{2}(1-x_5)$ and $2k_2=\frac{p}{2}(1+x_5)$.
The role of the radial coordinate in
the 2D QHE problem is played by $1-x_5$, which measures the
distance away from the origin of the droplet at $x_5=1$. In the 2D
case, the orbital angular momentum is simply a $U(1)$ phase
factor. In our case, the orbital angular momentum is a $SO(4)$
Casimir operator, whose eigenvalue is given $2k_1=\frac{p}{2}(1-x_5)$.
Therefore, just as in the 2D case, the distance away from the
center of the droplet directly determines the magnitude of the
orbital angular momentum. Because the confining potential can
be linearized near the edge of the droplet $1-x_5^F$, this
relationship translates into a massless relativistic dispersion relation.
Furthermore, as we shall see, the coupling to the iso-spin degrees of
freedom gives rise to particles with non-trivial helicity.

An edge excitation is created by removing a
particle (leaving behind a hole) inside the fermi latitude
$x_5^F$, with quantum numbers $(x_5^h;
k_1^h=\frac{p}{4}(1-x_5^h),k_{1z}^h; k_2^h=\frac{p}{4}(1+x_5^h),
k_{2z}^h)$, and creating a particle outside the fermi latitude,
with quantum numbers $(x_5^p; k_1^p=\frac{p}{4}(1-x_5^p),k_{1z}^p;
k_2^p=\frac{p}{4}(1+x_5^p), k_{2z}^p)$. This excitation can also
be specified by the quantum numbers $(\Delta x_5=x_5^h-x_5^p; T_1,
T_{1z}; T_2, T_{2z})$, where the total angular momenta
$T_{1i}=K_{1i}^h+K_{1i}^p$, $T_{2i}=K_{2i}^h+K_{2i}^p$,
$T_{1i}^2=T_1(T_1+1)$ and $T_{2i}^2=T_2(T_2+1)$ are the sums of
the $SU(2)\times SU(2)$ quantum numbers of the particle and the
hole. From the usual rules of the $SU(2)$ angular momentum
addition, we can determine the allowed values of the total angular
momenta $T_1=|k_1^p-k_1^h|,...,k_1^p+k_1^h$, and
$T_2=|k_2^p-k_2^h|,...,k_2^p+k_2^h$. Given $x_5^h$ and $x_5^p$
we obtain $\Delta x_5 = x_5^h-x_5^p=\frac{2}{p} n$, and the energy is given by
\begin{eqnarray}
E\approx \frac{\partial V}{\partial X_5} \Delta X_5 =
\frac{\partial V}{\partial X_5} \frac{2R}{p} n
\label{energy}
\end{eqnarray}

In the 2D QHE case, there is an unique way to combine the angular
momenta of a particle and a hole, therefore, the dispersion
relation has only one branch. In higher dimensions, a particle and
a hole can be bound or independent, giving rise to collective
and continuum branches of the spectrum. Mathematically, this
effect manifests itself in terms of the different ways of
combining the $SO(4)$ angular momenta of a particle and a hole.
Let us investigate the possibility of collective excitations
in the spectrum. In a non-interacting fermi system
with the usual form of the kinetic energy, $E={\bf p}^2/2M$, a particle
and a hole have a well defined relative momentum, but does not
have a well defined relative position, except in one spatial
dimension. Therefore, such a pair can only be ``bound" through an
attractive interaction. However, there are very special cases
where the pair can be bound for kinematic reason, without any
interactions. In one dimension, the kinetic energy is
approximately independent of the relative momentum, therefore, one
can superpose states with different relative momenta to obtain a
state with well defined relative position. The resulting state is
a bosonic collective mode. In our case, we find that the special
nature of the wave function in the lowest $SO(5)$ level leads to a
similar form of the kinematic binding. Basically, there is no
kinetic energy in the lowest $SO(5)$ level, a particle and a hole
can be locked into a well defined relative position without any
cost of the kinetic energy. In our case, these collective excitations
lie at the edge of the continuum states, and are characterized
by the total $SO(4)$ quantum numbers
$(T_1=|k_1^p-k_1^h|=\frac{n}{2},T_2=T_1+|\lambda|)$ and
$(T_1=T_2+|\lambda|,T_2=|k_2^p-k_2^h|=\frac{n}{2})$, where
$|\lambda|$ is a positive integer and $\lambda=0$ case is counted
only once. These states are formed by a macroscopic number of
contractions of the spinor wave functions (Eq. \ref{wavefunction})
of a particle and a hole, and it can be shown explicitly that
the wave function in the relative orbital and iso-spin coordinates
are gaussianly localized. In this sense, a particle and a hole
form a bound state, and represent collective excitations of the
system.

In the flat space limit, the $SO(4)$ symmetry group of $S^3$
reduces to the Euclidean group $E_3$ of the three dimensional flat
space. The Euclidean group has two Casimir operators, the
magnitude of the momentum operator $|{\bf p}|$ is determined by
either $T_1$ or $T_2$, which in our case gives  $|{\bf p}|= n/R$.
As the energy is given by Eq. \ref{energy}, the collective excitations
have a relativistic linear dispersion relation $E=c|{\bf p}|$,
with the speed of light given by
$c=\frac{\partial V}{\partial X_5} \frac{2R^2}{p}
=2 l_0^2 \frac{\partial V}{\partial X_5}$.
If we take for $l_0$ the Planck length
$l_P=1.6\times 10^{-35} m$,
we can estimate the potential energy gradient to be
$\frac{\partial V}{\partial X_5}\approx 7.7\times 10^{62} eVm^{-1}$.

The second Casimir operator of the Euclidean group is the helicity,
$\lambda={\bf J}\cdot {\bf p}/|{\bf p}|$, where ${\bf J}$ is the total
angular momentum of a particle. This quantity can be obtained from the
$SO(4)$ quantum numbers by $\lambda = T_1-T_2$\cite{talman}.
Therefore, the $(T_1=\frac{n}{2},T_2=T_1)$ state describe a
relativistic spinless particle obeying the massless Klein-Gordon
equation. The $(T_1=\frac{n}{2},T_2=T_1+1)$ and the
$(T_1=T_2+1,T_2=\frac{n}{2})$ states describe massless photon
states with left handed and right handed circular polarization.
The associated fields satisfy the Maxwell's equation.
The $(T_1=\frac{n}{2},T_2=T_1+2)$ and the
$(T_1=T_2+2,T_2=\frac{n}{2})$ states describe massless graviton
states with left handed and right handed circular polarization.
The associated fields satisfy the linearized Einstein equation.
In fact, we can proceed this way to find all massless relativistic
particles with higher spins. Here the time dimension is introduced
to the problem through the energy of the confining potential
(Eq. \ref{energy}), while the space dimension is introduced through
the Euclidean momentum. The relativistic dispersion
together with the helicity quantum numbers show that the
collective excitations form non-trivial representations of the
Lorentz group. The spins of these massless particles are derived from the
isospin degrees of freedom in the original Hamiltonian, and the
relativistic field equations have their roots in the original isospin-orbital
couplings.

So far we obtained only a non-interacting theory of relativistic
particles, in particular, the equation for the graviton is only
obtained to the linear order. Once we turn on interactions among
the different modes, the graviton would naturally couple to
the energy momentum tensor of other particles. It is known that
consistency requires the graviton to couple itself exactly according
to the full nonlinear Einstein equation\cite{feynman,weinberg}.
Therefore, it is likely that the interaction among the edge
modes in our model also contains the nonlinear effects of quantum
gravity.
On the other hand, the main problem with the current model seems to be the
``embarrassment of riches".
In order to define a problem with large degeneracy in the single particle
spectrum, one needs to take the limit of high representation of
the isospin. Therefore, each particle has a large number of internal
degrees of freedom. As a result, there are not only photons and gravitons
in the collective modes spectrum, there are also other massless
relativistic particles with higher spins.
However, the presence of massless higher spin states
may not lead to phenomenological contradictions. It is known
from the field theory that massless relativistic particles with spin
$s>2$ can not have covariant couplings to photons and gravitons\cite{pauli}.
Therefore, it is possible that they decouple in the long wave length limit.

{\bf Hall current and noncommutative geometry:}
So far, we have discussed only the quantum eigenvalue problem,
it is also instructive to discuss the classical Newtonian equation
of motion derived from the Hamiltonian $H+V(X_a)$, where $H$ is
given by Eq. \ref{H}. The classical degrees of freedom are the isospin
vector $I_i$, the position $X_a$ and the angular momentum $L_{ab}$,
and their equations of motion can be derived from their Poisson
bracket with the Hamiltonian. As we are interested in the equations
of motion in the lowest $SO(5)$ level, we can take the infinite mass
limit $M\rightarrow \infty$. In this limit, we obtain the following
equations of motion:
\begin{eqnarray}
\dot X_a = \frac{R^4}{I^2}\frac{\partial V}{\partial X_b} F^i_{ab} I_i \ \ \ , \ \ \
\dot I_i = \epsilon_{ijk} A_\mu^j \dot X_\mu I_k
\label{newton}
\end{eqnarray}
where the dot denotes the time derivative.
Just as in the lll problem, the momentum variables can be fully
eliminated. However, the price one needs to pay for this elimination
is that coordinates $[X_a,X_b]$ becomes non-commuting. In fact,
the projected Hamiltonian in the lowest $SO(5)$ level is simply
$V(X_a)$. If we assume the commutation relation
$[X_a,X_b]=\frac{R^4}{I^2}F_{ab}$, then the orbital part of Eq. (\ref{newton})
can be derived from the Poisson bracket of $X_a$ with $V(X_a)$.
If we expand around the north pole
$X_5=R$, we finally obtain the following commutation relation:
\begin{eqnarray}
[X_\mu,X_\nu]= 4 i l_0^2 \eta_{\mu\nu}^i \frac{I_i}{I}
\label{non-commutation}
\end{eqnarray}
This is the central equation underlying the algebraic structure of this
work. It shows that there is a fundamental limit, $l_0$, for the measurability
of the position of a particle.

The first equation in Eq. (\ref{newton}) determines the Hall current for a
given spin direction $J_\mu^i$ in terms of the gradient of the
potential $\eta^i_{\mu\nu}\partial V/\partial X_\nu$, giving a direct
generalization of the 2D Hall effect.
From the second equation in Eq. (\ref{newton}), we see that the spin
of a particle precesses around its orbital angular momentum
(which becomes linear momentum in the flat space limit) with a
definite sense.

{\bf Conclusion:}
At the conclusion of this work, we now know three different
spatial dimensions where quantum disordered liquids exist:
the one dimensional Luttinger liquid, the two
dimensional quantum Hall liquid, and the four dimensional
generalization found in this work.
We can ask what makes these dimensions special.
There is a special mathematical
property which singles out these spatial dimensions.
One, two and four dimensional
spaces have the unique mathematical property that
boundaries of these spaces are isomorphic to mathematical groups,
namely the groups $Z_2$, $U(1)$ and $SU(2)$. No other spaces have
this property. It is this deep connection between the algebra
and the geometry which makes the construction of non-trivial
quantum ground states possible. Other related mathematical connections are
reviewed and summarized in ref. \cite{holonomy}.
The 4D generalization of the QHE offers an ideal theoretical
laboratory to study interplay between quantum correlations and
dimensionality in strongly correlated systems.
It would be interesting to study our quantum wave functions on
four dimensional manifolds with non-trivial topology, and
investigate if different topologies of four manifolds
correspond to degeneracies of our many body ground states.
The quantum plateau transition in the 2D QHE is
still an unsolved problem, one could naturally ask if the plateau
transition in four dimensions can be understood better because
of the higher dimensionality. In 2D QHE,
quasi-particles have both anyonic and exclusion
statistics. The former can not exist in four dimensions, the
question is whether quasi-particles in our theory would obey
exclusion statistics in the sense of Haldane.
To address these questions, it is important to construct a
field theory description of the 4D quantum Hall liquid, in
analogy with the Chern-Simons-Landau-Ginzburg theory of the
QHE.

In this work we investigated the possibility of modeling
relativistic elementary particles as collective boundary
excitations of the 4D quantum Hall liquid.
Similar connections between condensed matter and
particle physics have been explored
before\cite{bj,nielsen,volovik,chapline,lenny}.
There are important aspects unique to the current problem. The
single particle states are hugely degenerate, which enables the
limit of zero inertia mass $M\rightarrow 0$ and completely
removes the non-relativistic dispersion effects. This limit is hard to
take in usual condensed matter systems. The single particle states
also have a strong gauge coupling between iso-spin and orbital degrees of freedom,
which is ultimately responsible for the emergence of the relativistic
helicity of the collective modes. This type of coupling is not
present in usual condensed matter systems. The vanishing of the kinetic energy
in the lowest $SO(5)$ levels enables binding of a particle and a hole
into a point-like collective mode.
The most remarkable mathematical structure is the non-commutative
geometry (Eq. \ref{non-commutation}), which expresses a $SU(2)$
co-cycle structure of the magnetic translation.
Although progress reported in this work is still very
limited, we hope that this framework can stimulate investigations on
the deep connection between condensed matter and elementary
particle physics.

\newpage
{\bf Note added}

Since the submission of this paper, some questions have been raised.
We would like to make the following qualifying statement to the paper,
as to avoid misinterpretation of results:

In usual non-interacting fermion systems, for
a given center of mass momentum, a particle hole excitation can
either have a well defined energy but no well defined relative
position, (in other words, they can not be created by local
operators), or they can have a well defined relative position
but no well defined energy.

Because of the non-commutative geometry given in equation (14),
the collective modes, or better phrased as extremal dipole states,
studied in this paper have both well
defined energy and well defined relative position, for a fixed
center of mass momentum, even though they are composed out of
non-interacting fermions. They can be created by local bosonic
operators and and these local bosonic operators obey relativistic
equation of motion, with well defined dispersion relation.

These excitations can also be interpreted as hydrodynamical waves with
different spins on the surface of the 4D QHE droplet. Since the
fermionic particles carry a high spin $I$, there are many different
branches of hydrodynamical modes, corresponding to bosonic excitations
of different helicities.

Why do these hydrodynamical modes come with both helicities, and are not
chiral, as in the 1D case? Consider our 4D droplet, the scalar density wave
where fermions with all different iso-spin components are compressed in the same way.
This mode has to trivially obey the Klein-Gordon equation. There is
no concept of chiral bosons in 3+1 dimensions. Since the scalar mode is
symmetric, it is plausible that all other modes come with both helicities.

Incoherent part of the fermionic spectrum are not relativistic.
Currently we are investigating mechanisms by which the incoherent
part of fermionic spectrum can be gapped due to interactions, leaving
collective modes unaffected. One known example of such behavior
is superconductivity. In this system, other mechanism may be possible as
well. In this case, the theory would be fully relativistic in the low energy
sector.

\end{document}